\pdfoutput=1
\documentclass{JINST}
\usepackage{graphicx}
\usepackage{amssymb,amsmath}
\usepackage{epstopdf}
\usepackage{upgreek}

\newcommand{\about}{\mathord{\sim}}

\title{Measurements of wavelength-dependent double photoelectron emission from single photons in VUV-sensitive photomultiplier tubes}

\author{C.H.~Faham$^{a}$, V.M.~Gehman$^{a}$\thanks{Corresponding author: vmgehman@lbl.gov}, A.~Currie$^{a,b}$, A.~Dobi$^{a}$, P.~Sorensen$^{a}$~ and R.J.~Gaitskell$^{c}$\\
\llap{$^a$}{Lawrence Berkeley National Laboratory, 1 Cyclotron Rd., Berkeley CA 94720, USA}\\
\llap{$^b$}{Imperial College London, High Energy Physics, Blackett Laboratory, London SW7 2BZ, UK}\\
\llap{$^c$}{Brown University, Dept. of Physics, 182 Hope St., Providence RI 02912, USA}}

\abstract{Measurements of double photoelectron emission (DPE) probabilities as a function of wavelength are reported for Hamamatsu R8778, R8520, and R11410 VUV-sensitive photomultiplier tubes (PMTs). In DPE, a single photon strikes the PMT photocathode and produces two photoelectrons instead of a single one. It was found that the fraction of detected photons that result in DPE emission is a function of the incident photon wavelength, and manifests itself below $\sim$250~nm. For the xenon scintillation wavelength of 175~nm, a DPE probability of 18--24\% was measured depending on the tube and measurement method. This wavelength-dependent single photon response has implications for the energy calibration and photon counting of current and future liquid xenon detectors such as LUX, LZ, XENON100/1T, Panda-X and XMASS.}

\begin{document}

\section{Introduction}\label{sec:Intro}
Photomultiplier tubes (PMTs) are currently one of the most widely used photodetectors due to their large collection area and high gain. In radiation detectors using xenon as an active medium, vacuum ultraviolet (VUV)-sensitive PMTs are used to detect the 175~nm xenon scintillation light \cite{UCLAPaper, CarlosThesis}. For photon-counting and calorimetry applications, the single-photon response of PMTs must be well characterized to understand the overall detector behavior.

The most important contribution to the PMT photon detection efficiency is the quantum efficiency (QE), which is defined as the ratio of electrons produced in the photocathode (referred to as photoelectrons, or phe) divided by the number of incident photons \cite{HamamatsuManual}. QE is often incorrectly treated as the probability of an incident photon to produce a photoelectron, but if more than one photoelectron is produced by a single photon, this equivalence is no longer valid.  QE is a function of the incoming photon wavelength due to a number of competing factors, including the PMT window transmission probability and the photon absorption length in the photocathode. QE is typically limited to $\about30\%$ at 400~nm, with decreasing sensitivity at lower wavelengths. For PMTs specially designed to be sensitive to the VUV band, the QE shows a marked enhancement below 200~nm and above the transmission cutoff of the synthetic silica window at 160--170~nm\cite{UCLAPaper, CarlosThesis, Dorenbos:1993}.  The present work was prompted by observed differences in the PMT pulse height and area spectra for single photons of xenon scintillation light and blue LED light in LUX\cite{luxSD, luxPRL}.  Specifically, a feature like the one discussed in Section \ref{sec:Results} of this document, led us to start investigating not only the efficiency, but also the shape of the single photon response of LUX PMTs as a function of incoming wavelength.

The process of emitting two photoelectrons from the photocathode from a single incoming photon, referred to as double photoelectron emission (DPE), has been studied before in the context of the photoelectric effect on metallic surfaces \cite{DPE1, DPE2, DPE3}. PMT photocathodes are expected to exhibit a similar behavior, provided that the incoming photon is sufficiently energetic (greater than twice the photocathode work function). We have found in this study that for wavelengths lower than $\about 250$~nm, the DPE probability is non-negligible for all three VUV-sensitive Hamamatsu R11410 (3-inch circular photocathode), Hamamatsu R8778 (2-inch circular) and R8520 (1-inch square). The R8520 PMT was used in the XENON10, XENON100, Panda-X (top array) and LZ (xenon veto region) experiments. The R8778 was utilized in the LUX experiment, while a variant (R10789 hexagonal head) was used in the XMASS experiment. Lastly, the larger R11410 has been used in the Panda-X experiment (bottom array) and will be used by next-generation dark matter experiments, such as LZ, XENON1T/NT, and RED, among others.

\section{Experimental Setup}\label{sec:Equipment}
\begin{figure}[h]
\begin{center}
\includegraphics[width=12cm]{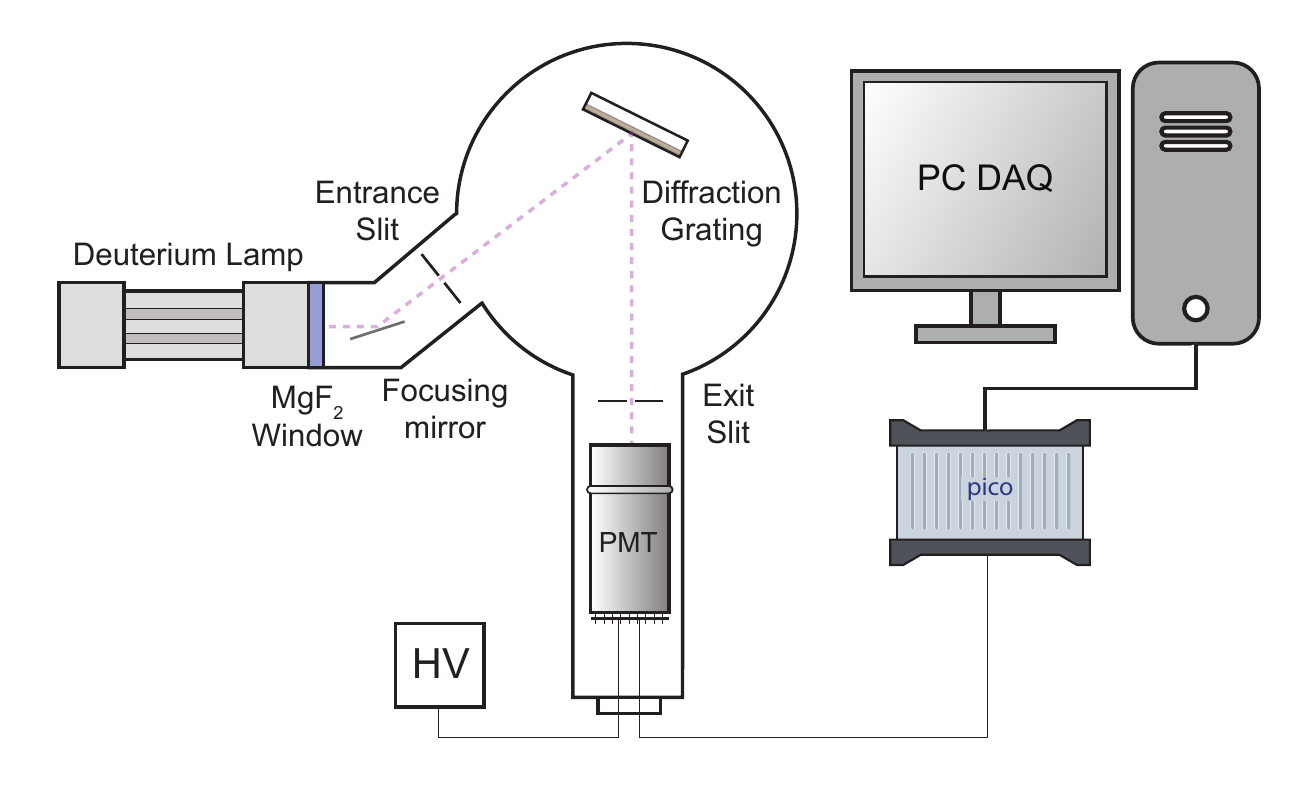}
\caption{Experimental setup for the DPE measurement.}
\label{fig:experimental_setup}
\end{center}
\end{figure}

The experimental setup used for all measurements is shown schematically in Figure~\ref{fig:experimental_setup} and is similar to the one used in References \cite{Gehman:2011, Gehman:2013A, Gehman:2013B}.  A Model 632 Deuterium Light source, from McPherson, Inc.\cite{McPherson} created a broad spectrum extending from 110--950~nm with the window facing into a vacuum space pumped down to below $10^{-4}$~Torr.  A focusing mirror imaged the lamp window on a holographic diffraction grating inside a Model 234/302VM 0.2-Meter Monochromator, also from McPherson, Inc.  The grating separated the incoming light in reflected angle by wavelength and could be rotated to select a specific wavelength through the exit of the monochromator.  The monochromator body has adjustable slits (5~mm tall by 0--3~mm wide) at its entrance and exit.  To first order, the entrance slit only adjusts the intensity of the light exiting the monochromator, while the exit slit also adjusts the width in wavelength of the monochromator output.  A narrower exit slit samples a smaller range of reflected angles from the grating and therefore results in a narrower wavelength distribution.  For the duration of this study, the exit slit was set to 0.5~mm and the entrance ranged from 0.2--0.5~mm to regularize the single photon rate as measured by counting peaks above threshold in the data stream.  In this configuration the, full-width, half-maximum of the light source output was measured to be 4.4~nm between 200 and 300~nm with an Ocean Optics QE65000 UV/Vis spectrometer (which is only sensitive above 200 nm). Second-order (half-wavelength) peaks are present in the monochromator setup. Because the synthetic silica windows in all three PMT models are perfectly opaque to light below 155~nm, the maximum wavelength used in this study was 300~nm, which ensured no second-order peaks were detected by any of the PMTs. All of the PMT measurements were performed at room temperature.

\begin{table}[htp]
\caption{Bias and acquisition settings for each of the 3 tested PMTs.}
\begin{center}
\begin{tabular}{l c c c c} 
              & Mean single phe  & \\
PMT type      & height [mV]      & DAQ threshold [mV]\\
\hline
1-inch R8520  &  9.2             & 0.75 (0.082 single phe)\\
2-inch R8778  & 36.8             & 1.55 (0.042 single phe)\\
3-inch R11410 & 21.2             & 1.40 (0.066 single phe)\\
\end{tabular}
\end{center}
\label{table:pmt_settings}
\end{table}%

\begin{figure}[h]
\begin{center}
\includegraphics[width=12cm]{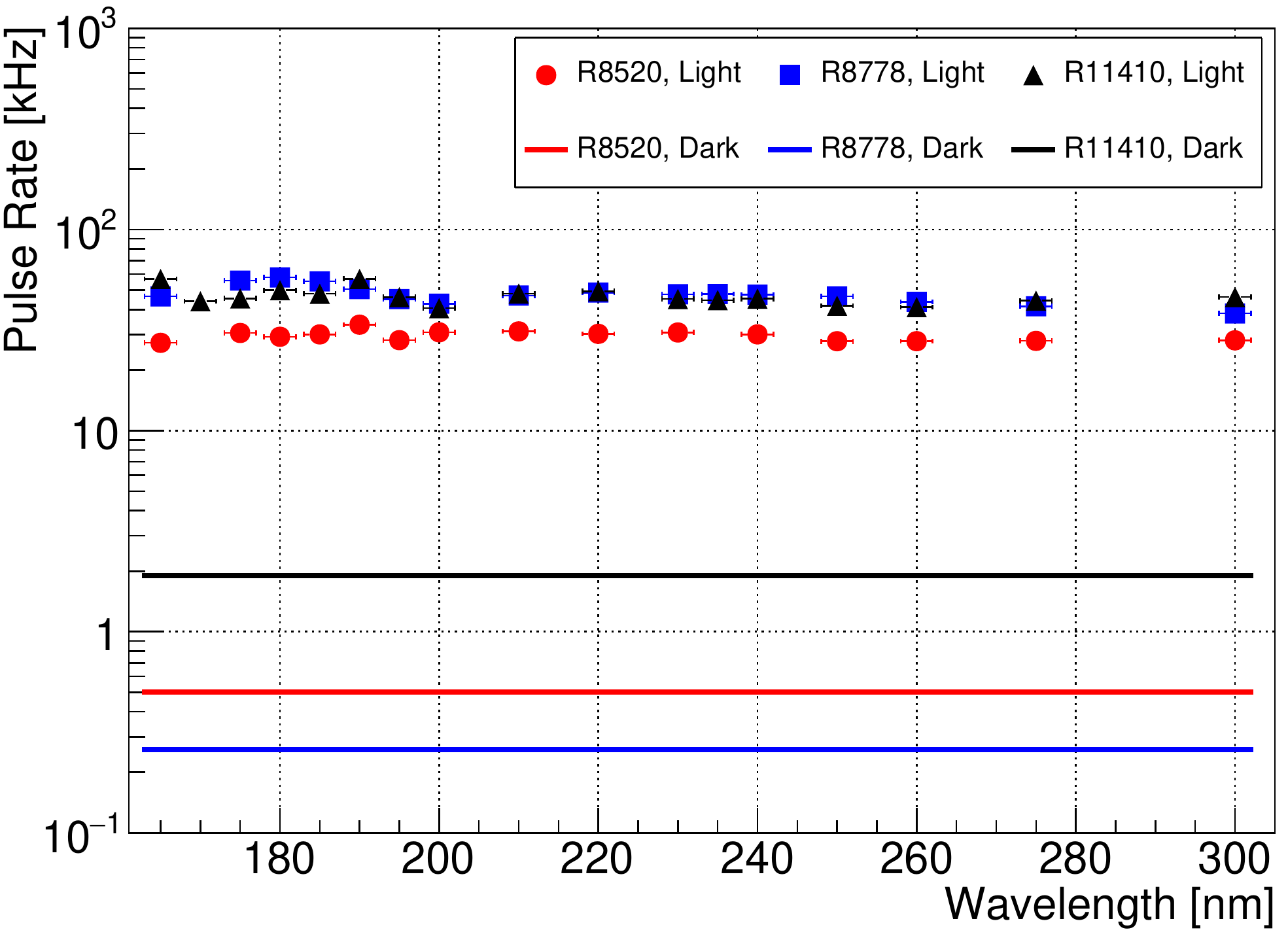} 
\caption{Rate of single photon and dark pulses above DAQ threshold during data acquisition for all wavelengths and PMT models (top panel).  Bottom panel shows the same photon rate data as the top, but plotted on a linear scale for easier comparison.}
\label{fig:GlobalRate}
\end{center}
\end{figure}

The PMTs examined in this study were mounted in black Delrin$\circledR$  holders to provide electrical insulation from their surrounding grounded flange, as well as to minimize reflected light.  PMT high voltage bias and signal were fed in and read out through coaxial feedthroughs on the vacuum.  PMT signals were digitized by a PicoScope 5000 series oscilloscope.  The PicoScope was triggered every 10 ms by an external pulse generator, and read out using a custom MATLAB \cite{MATLAB} data acquisition (DAQ) code. When the PicoScope received a trigger signal, the DAQ code read in a 5-ms long buffer, which was sampled at 500~MS/s (2~ns samples), and pulses above threshold were selected.  The pulse threshold was set depending on noise levels and gain for each PMT, and ranged from 0.75--1.55 mV.  Threshold and single PE peak values are reported in Table \ref{table:pmt_settings}.  For each pulse above threshold, a waveform of 150 samples (300 ns) centered around the pulse maximum was selected and was then baseline-corrected by subtracting the average of the first 10 samples. This pulse selection was performed in order of descending pulse size and the selection finished when there were no pulses left in the 5~ms buffer with a height above the set threshold value. Each acquisition (one per wavelength) was saved to a .mat file with the pulse waveform data and acquisition metadata. A total of 1~million pulses were collected at each wavelength for each PMT. The rate at which pulses were collected is plotted in Figure~\ref{fig:GlobalRate} and was calculated by dividing the number of observed pulses by the total amount of time searched.  At these rates, the pileup fraction, {\it i.e.}~the fraction of 300~ns windows containing $>$1 pulse, ranged from 0.9\% to 1.5\%.

\section{Results}\label{sec:Results}
\begin{figure}[h]
\begin{center}
\includegraphics[width=12cm]{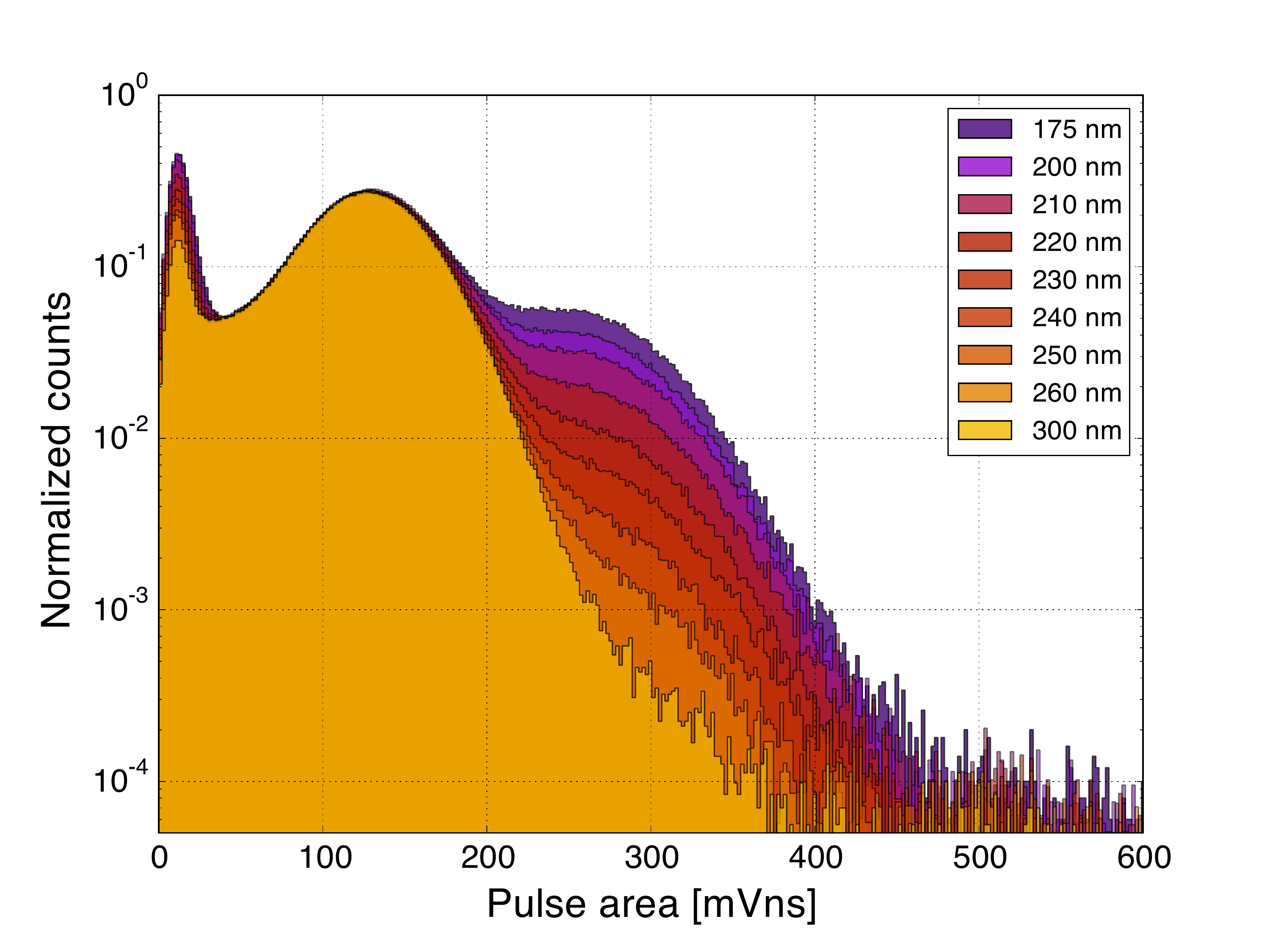} 
\caption{Superposition of pulse area spectra for the R11410 PMT for different wavelengths. Each spectrum is normalized by the integral in the region between 50--120~mV~ns in order to show the effect more clearly.}
\label{fig:all_wavelengths_spectra}
\end{center}
\end{figure}

The waveforms were loaded and analyzed in Python using SciPy \cite{SciPy} and pyROOT \cite{pyROOT}.  Single photon response spectra for each PMT at each wavelength were created by histogramming the waveform area.  We show the qualitative variation of these pulse area histograms for the R11410 as a function of wavelength in Figure \ref{fig:all_wavelengths_spectra}.
The spectra in Figure~\ref{fig:all_wavelengths_spectra} were normalized by the integral in the region between 50-120~mVns to better illustrate the increase of the amplitude of the double photoelectron peak with decreasing photon wavelength.  

\begin{figure}[h]
\begin{center}
\includegraphics[width=12cm]{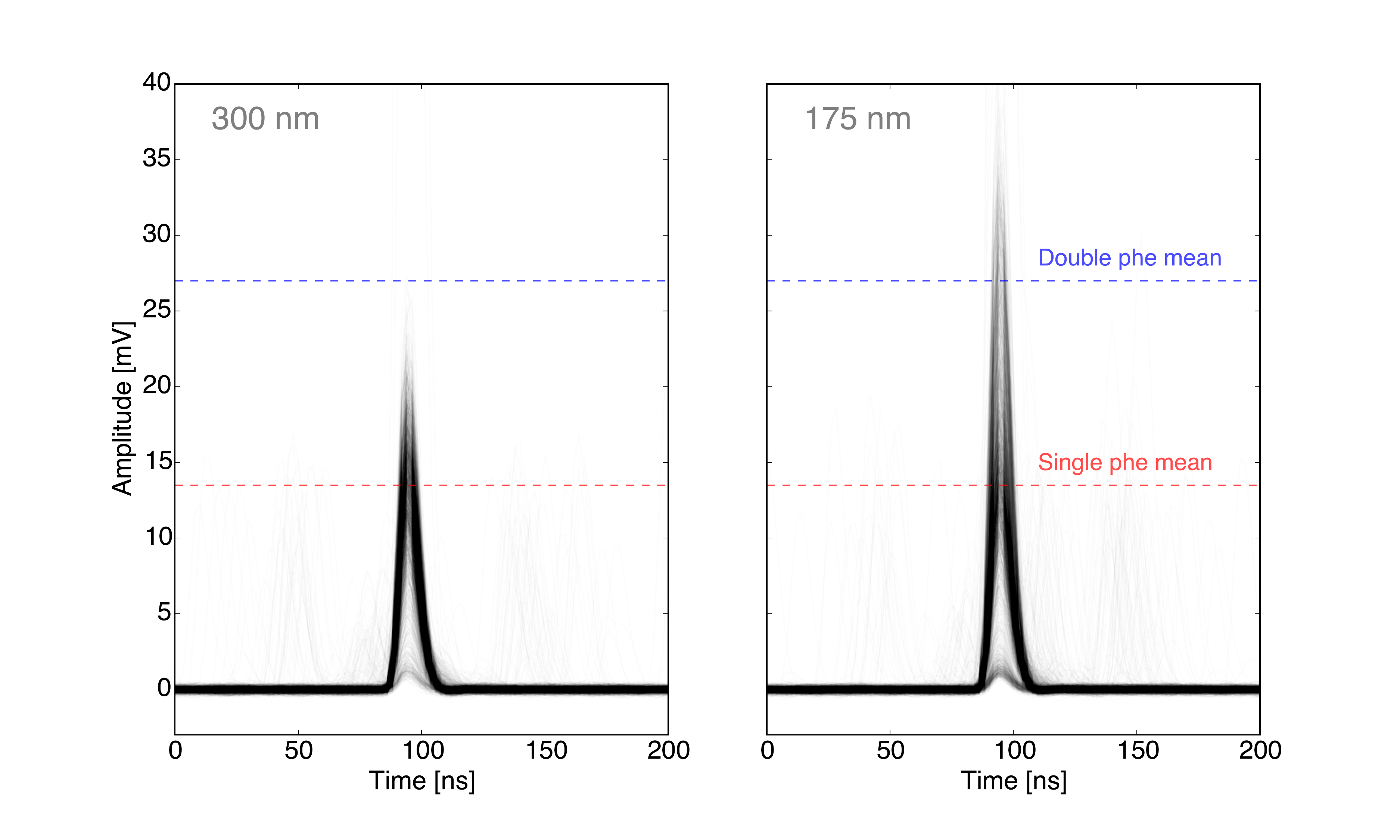} 
\caption{First 2000 pulses at 300~nm (left) and 175~nm (right) for the R11410 PMT acquisition. The pulse rates were 46.1~kHz and 45.4~kHz, respectively. All acquired waveforms are shown, but pulse quality cuts to remove pileup pulses were applied for the analysis. Direct photon hits on the PMT first dynode can be seen at approximately 1.5~mV in height. The rates of both first dynode photoelectron emission and double photoelectrons, relative to that of single photoelectrons, are higher at lower wavelengths.}
\label{fig:overlaid_pulse_traces}
\end{center}
\end{figure}

We further illustrate this difference in Figure \ref{fig:overlaid_pulse_traces}, where we have plotted an overlay of the first 2000 pulses collected at both 300~nm (left) and 175~nm (right) for the R11410. The 175~nm waveforms clearly have a higher amplitude and effectively the same width compared to the 300~nm waveforms.  The photon rate recorded by our DAQ was nearly identical: 46.1~kHz at 300~nm and 45.4~kHz at 175~nm.  The central region of \ref{fig:overlaid_pulse_traces} is centered at just less than 100~ns.  Note that there are a few dozen pulses that fall outside of the central time.  This is consistent with our estimate of the pileup rate, and these pulses were removed from our subsequent analysis with a simple pulse quality cut to remove events with more than one significant local maximum.

Another feature visible in both Figures \ref{fig:all_wavelengths_spectra} and \ref{fig:overlaid_pulse_traces} is a population of pulses arising from incident photons which are not absorbed by the photocathode, but create a photoelectron on the first dynode instead.  These events populate the peak in the pulse area histogram centered around 20 mV~ns in Figure \ref{fig:all_wavelengths_spectra} and the persistent trace visible at 1.5 mV in amplitude in Figure \ref{fig:overlaid_pulse_traces}.  We verified this by running with the photocathode and first dynode pins on the PMT base shorted together and the overall bias adjusted to maintain the same field between the rest of the dynodes as in regular operating conditions.  In this mode, the efficiency for collecting electrons from the photocathode is very small, while the efficiency for collecting electrons from the first dynode remains unchanged.  When running in this shorted photocathode configuration, the mean and width of this 20 mV~ns distribution remain unchanged while the single and double phe distribution vanish.  The first dynode photoelectrons should in principle, exhibit a similar DPE behavior as those from the photocathode.  This is an interesting topic for future study, but lies outside the scope of this article.

The pulse area histograms were used to calculate the fraction of PMT pulses exhibiting double photoelectron emission as a function of wavelength. Two different methods were employed: 

\begin{enumerate}
\item Perform a simultaneous single and double photoelectron distribution fit for each spectrum, and use the fitted areas to estimate the DPE fraction.
\item Calculate the shift in the mean pulse area, which should directly correspond to the DPE fraction.
\end{enumerate}

\begin{figure}[h]
\begin{center}
\includegraphics[width=12cm]{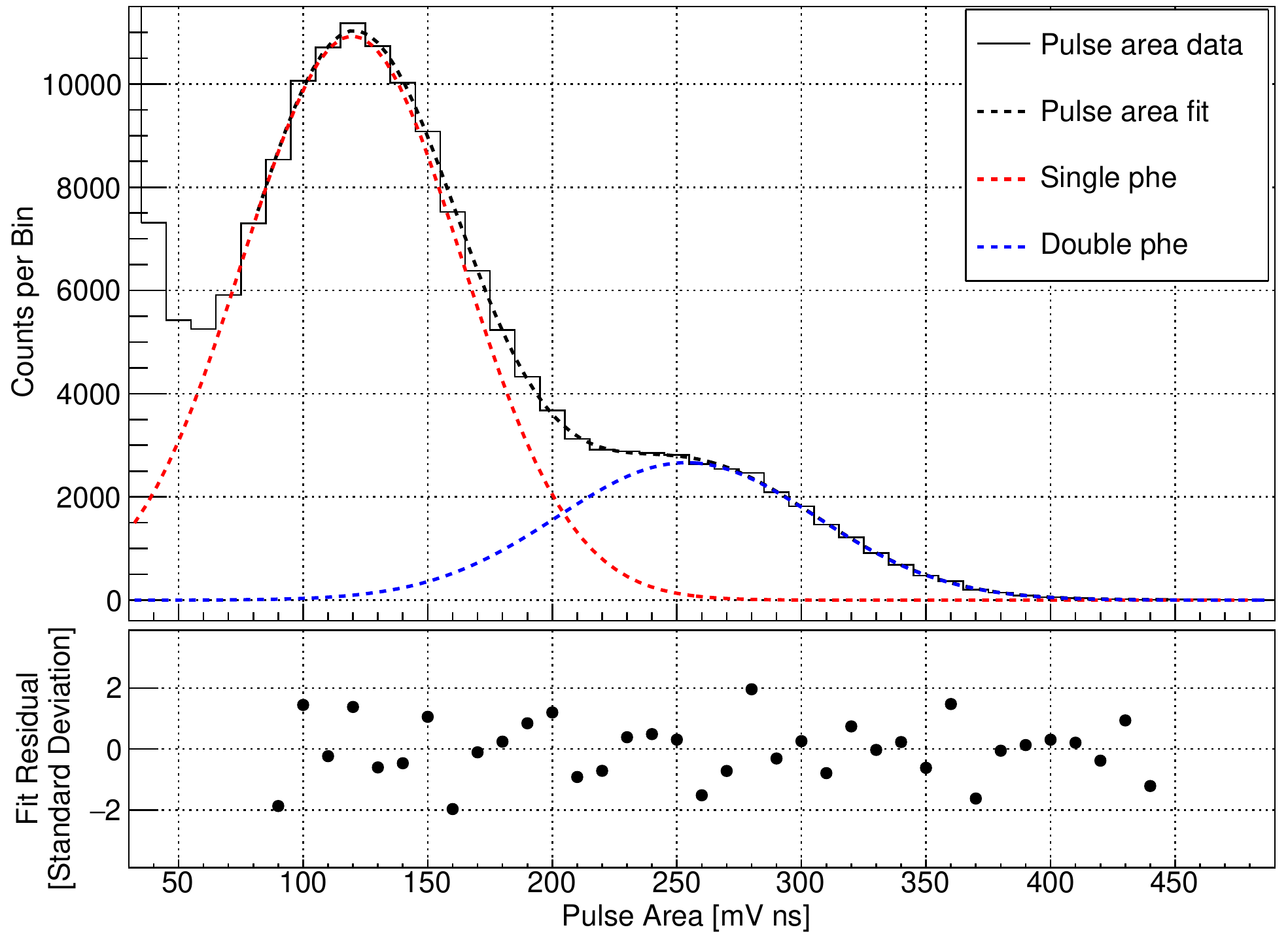} 
\caption{Fit to the single, and double photoelectron spectrum (top panel) and fit residuals (bottom panel) for the R11410 at 175~nm.  Residuals are normalized to the bin-by-bin uncertainty (assuming simple Poisson counting statistics) and are plotted over the fit range.  The DPE fraction for this fit to the spectrum is $22.5 \pm 1.0$\%.}
\label{fig:2phe_peak_extraction}
\end{center}
\end{figure}

For the first method, our fit model was a pair of Gaussian functions\footnote{While PMT pulse area distributions can have significant deviations from Gaussian in the limit of sufficiently large statistics, the agreement between this model and this data is excellent.  The Gaussian-based model used in this article is much easier to interpret than others, such as Polya or scaled Poisson distributions because it fits the mean and width of each peak directly.} plus a constant background offset:
\begin{equation}
F(x) = N_{sphe} e^{-\left (\frac{x - \mu_{sphe}}{2\sigma_{sphe}}\right )^{2}} + N_{dphe} e^{-\left (\frac{x - \mu_{dphe}}{2\sigma_{dphe}}\right )^{2}} + C,
\end{equation}
where $N_{sphe}$ and $N_{dphe}$ are the normalizations, $\mu_{sphe}$ and $\mu_{dphe}$ are the mean values, $\sigma_{sphe}$ and $\sigma_{dphe}$ are the widths for the single and double photoelectron peaks, and $C$ is the constant background term.  The initial guesses for each fit come from a Gaussian fit to the region very near the single phe peak in the 300~nm spectrum, which should have the smallest DPE fraction for the acquired data.  The double phe peak parameter guesses are then calculated under the assumption that the double phe population is defined by sampling the single phe population twice, {\it i.e.} $\mu_{dphe} = 2\ \mu_{sphe}$ and $\sigma_{dphe} = \sqrt{2}\ \sigma_{sphe}$.  Both means and widths were then allowed to float within $\pm$5\% of the initial guess value.  The DPE fraction for each spectrum is the area of the double phe distribution divided by the sum of the area of the single plus the double photoelectron distributions.  We show an example of a fit to the R11410 pulse area spectrum at 175~nm in Figure \ref{fig:2phe_peak_extraction}, for which the 
DPE fraction 
was calculated to be $22.5 \pm 1.0$\%.

To calculate the DPE fraction using the shift in the mean of the pulse area spectrum, we truncate the low side of the spectrum at the valley between the single phe peak and the first dynode and pedestal structure, {\it e.g.} around 50~mVns for the spectrum in Figure \ref{fig:2phe_peak_extraction}.  We then simply take the bin-weighted mean of each pulse area spectrum.  The mean shift is calculated relative to the 300~nm spectrum by the formula:
\begin{equation}
\mbox{Mean Shift}(\lambda) = \frac{\mu_{\mathrm{spec}}(\lambda) - \mu_{\mathrm{spec}}(\lambda = 300 \mbox{ nm})}{\mu_{\mathrm{spec}}(\lambda = 300 \mbox{ nm})},
\end{equation}
where $\mu_{\mathrm{spec}}(\lambda)$ is the mean of the pulse area spectrum at wavelength $\lambda$.  This quantity is therefore identically zero by construction at 300~nm, and allows for a simple, model-independent estimate of the DPE fraction versus wavelength.

\begin{figure}[h]
\begin{center}
\includegraphics[width=15cm]{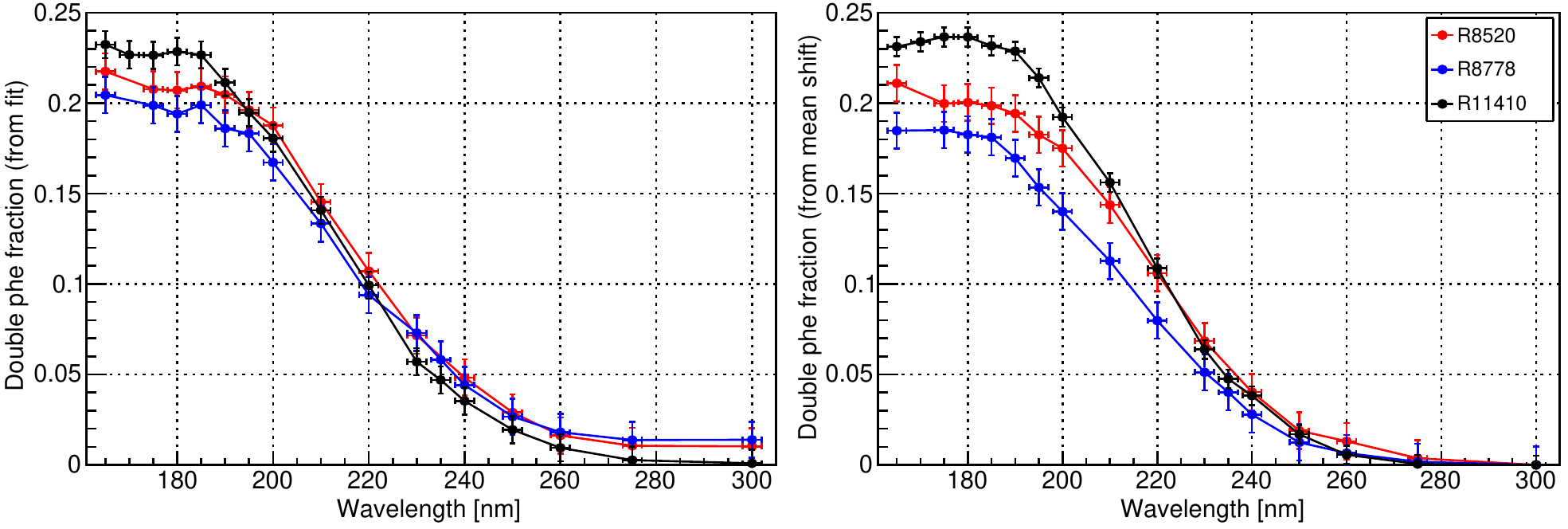}
\caption{DPE probability for all three PMTs tested as a function of incident photon wavelength using the fit (left) and mean shift (right) methods.}
\label{fig:dpe_probability_vs_wavelength}
\end{center}
\end{figure}

We plot the DPE fraction using both the fit (in the left panel) and mean shift (in the right panel) methods for each PMT in Figure \ref{fig:dpe_probability_vs_wavelength}.

Uncertainties in the traces of Figure \ref{fig:dpe_probability_vs_wavelength} are a quadrature sum of two effects.  One is the intrinsic uncertainty from the process of calculating the DPE fraction.  This incorporates the fit uncertainties in the left panel and the counting statistics uncertainties for the right.  The second one is the systematic uncertainty from the range over which each method was carried out, and this is the dominant contribution to the overall uncertainty.  For the fit, this is just the uncertainty arising from changes to the fit range.  For the mean shift, this is the uncertainty arising from adjusting the range over which the histogram mean was calculated.  Both methods agree in the overall DPE fracion vs. wavelength trend, and point to a DPE fraction that turns on between 250 and 260~nm, climbs steadily with decreasing wavelength, and reaches its maximum value of approximately 20\% near 185~nm.

\section{Implications for QE Measurement and Photon Counting in Xenon Experiments}\label{sec:Implications}

PMTs are commonly used to detect photons in the visible band, and visible light sources are more readily available for bench-top measurements.  This means nearly all PMT characteristics are measured at visible wavelengths, usually with blue (400--450~nm) light emitting diodes (LEDs). The fact that a single photon can cause more than one photoelectron to be emitted from a photocathode at VUV wavelengths makes it important to measure the relevant PMT characteristics in the VUV. It is therefore important to distinguish two different figures of merit in place of a single quantum efficiency $\eta$. Let $\eta_{\upmu}(\lambda)$ be the ratio of photoelectrons emitted from the photocathode to photons incident,
and let $\eta_{\mathrm{p}}(\lambda)$ be the fraction of incident photons generating one or more photolectrons. The former is the mean response and the latter is roughly (ignoring first dynode hits for the moment) the probability of a detectable response; in the regime where the emission of three or more photoelectrons is negligible, they are related by:
\begin{equation}
\eta_{\upmu}(\lambda) = (1+\textrm{DPE fraction})\, \eta_{\mathrm{p}}(\lambda) .\label{qe_defs}
\end{equation}
\noindent The distinction between these two efficiencies together with their wavelength dependence have several implications for experiments which measure VUV scintillation light with an array of single-phe sensitive PMTs.

The mean response $\eta_{\upmu}(\lambda)$ can be measured via the photocathode current under illumination by a calibrated light intensity, and is usually quoted in PMT spec sheets. However, it is $\eta_{\mathrm{p}}(\lambda)$ that often determines a detector's energy threshold via the probability for a given primary scintillation signal to meet an $n$-PMT coincidence requirement. Simulations which use measurements of $\eta_{\upmu}(\lambda)$ as if they were probabilities for detection will over-estimate efficiencies of this type whenever the DPE fraction is greater than zero.

Away from threshold, DPE affects both resolution and the calibration of absolute yields of quanta. For a given mean number of photoelectrons, the resolution of short-wavelength signals is worsened, with respect to what would be expected from longer-wavelength calibration, by both the smaller underlying binomial detection probability and the higher fractional standard deviation of the single-photon response. We suggest that $\eta_{\mathrm{p}}(\lambda)$, calculated for instance via equation \ref{qe_defs}, be quoted in the specification of VUV-sensitive photomultipliers to enable more accurate modeling of thresholds and resolution. Calibration of the single photon response using VUV signal light, as in \cite{Neves2009}, instead of longer wavelengths allows pulse areas to be interpreted directly as a number of detected signal photons without separate measurement of the DPE fraction.  Simple energy scale calibrations that derive from the comparison of signals summed over all PMTs to mono-energetic calibration lines from known radioactive sources (such lines are typically much higher energy than the threshold) should be largely insensitive to the DPE fraction because it is in both the numerator and denominator of that energy calibration ratio.  Energy calibrations which depend on a model of detector resolution, especially those using continuous spectra, must however account for the DPE fraction.

The relatively large variance of the VUV photon pulse area distribution, due to DPE, argues for counting integer photons in waveforms, in addition to integrating pulse areas, to optimize resolution of small scintillation signals. A further advantage is that first-dynode photoelectrons could be used to boost efficiency in the same few-photon regime. The first-dynode peak seen in Figure~\ref{fig:all_wavelengths_spectra} contributes approximately 20\% of the photons above threshold at 175~nm. Further study is needed of how much additional light collection can thus be gained in full detectors: the present results are for normally incident light in vacuum and the probability of transmission through the window and subsequent absorption on the first dynode depends on the surrounding material index of refraction and incident photon trajectory; additionally, dynode voltages could be tuned to raise more of the first dynode signal above the photon-counting threshold.

\section{Summary and Future Work}\label{sec:Summary}
This study addresses a number of interesting questions that impact numerous physics programs that use xenon as a radiation detection medium.  The normalization of PMT signals to count photoelectrons is core to energy calibration and acceptance characterization, and this article addresses a number of important wavelength dependent issues.  The magnitude of the effect shown in Figure \ref{fig:dpe_probability_vs_wavelength} is similar to that seen in LUX single-photon spectra and may account for the short-wavelength enhancement of $\eta_{\upmu}(\lambda)$ in VUV-sensitive PMTs \cite{UCLAPaper}.

There are a number of follow-up measurements that could be done to further investigate the emission of two photoelectrons into the VUV.  First, there is a 10--15\% relative variation in the DPE fraction at the shortest wavelengths to which these PMTs are sensitive.  Because only one PMT of each model was tested for this study, it remains unclear whether this corresponds to variation within or between PMT models.  A study involving a larger number of each device would help to address this question.  Furthermore, most xenon-filled detectors operate predominately in the liquid phase, meaning that the PMTs must operate at cryogenic temperatures (165~K).  The tests for this study were performed at room temperature, so a future study examining temperature dependence of the DPE fraction would be very informative.  Another subject for future study would be to check if the calibrated photodiodes used for absolute QE measurements exhibit this same DPE effect, and how that affects the wavelength-dependent QE measurements that the manufacturer provides.  

\section{Acknowledgements}
The authors of this paper would like to thank Murdock Gilchriese, Kevin Lesko, Matthew Szydagis, and the entire LUX collaboration for their support and constructive feedback, as well as Keith Rielage and Los Alamos National Laboratory for the loan of several pieces of critical spectroscopy equipment.  This work was supported by the Director, Office of Science, of the U.S. Department of Energy under Contract Numbers DE-AC02-05CH11231 and DE-SC0010010.

\clearpage
\pagebreak


\begin{thebibliography}{999}
\bibitem{UCLAPaper} A. Lyashenko, {\it et al}. {\it Journal of Instrumentation} {\bf 9} (2014) P11021. arxiv:1410.3890.  DOI:doi:10.1088/1748-0221/9/11/P11021.
\bibitem{CarlosThesis} C.~H.~Faham. ``Prototype, Surface Commissioning and Photomultiplier Tube Characterization for the Large Underground Xenon (LUX) Direct Dark Matter Search Experiment." Ph.D Thesis, Brown University (2013).
\bibitem{HamamatsuManual} Hamamatsu Photonics K.K., ``Photomultiplier Tubes: Basics and Applications," Edition 3a (2007).
\bibitem{Dorenbos:1993} P. Dorenbos, {\it et al}. {\it Nucl. Instrum. Meth. A}, {\bf 325}, 367 (1993), DOI:10.1016/0168-9002(93)91040-T.
\bibitem{luxSD} D.S. Akerib et al. (LUX Collaboration).  ``Updated WIMP scattering limits from the first search run of the LUX experiment at the Sanford Underground Research Facility.''  In preparation.
\bibitem{luxPRL} D.S. Akerib, et al. (LUX Collaboration).  ``First Results from the LUX Dark Matter Experiment at the Sanford Underground Research Facility'' {\it Phys. Rev. Lett.} {\bf 112}, (2014) 091303.  DOI:10.1103/PhysRevLett.112.091303.  arXiv:1310.8214 [astro-ph.CO]
\bibitem{DPE1} Wolfgang Schattke, Michel A. Van Hove.  Solid-State Photoemission and Related Methods: Theory and Experiment.  John Wiley \& Sons, 2008.
\bibitem{DPE2} N. Fominykh, {\it et al}.  ``Theory of two-electron photoemission from surfaces.'' {\it Solid State Communications}.  {\bf 113} (2000) 665.
\bibitem{DPE3}Y. Pavlyukh, {\it et al}.  ``Single- or double-electron emission within the Keldysh nonequilibrium Green's function and Feshbach projection operator techniques.''  {\it Phys. Rev. B} {\bf 91} (2015) 155116.  DOI:10.1103/PhysRevB.91.155116.
\bibitem{Gehman:2011}  V.M.~Gehman, {\it et al}.  ``Fluorescence Efficiency and Visible Re-emission Spectrum of Tetraphenyl Butadiene Films at Extreme Ultraviolet Wavelengths.''  {\it Nucl. Instrum. Meth. A}, {\bf 654}, 116 (2011).  arXiv:1104.3259 [astro-ph.IM].  DOI:10.1016/j.nima.2011.06.088.
\bibitem{Gehman:2013A}  V.M.~Gehman, {\it et al}.  ``Characterization of protonated and deuterated Tetra-Phenyl Butadiene Film in a Polystyrene Matrix ''  {\it J.\ Inst.} {\bf 8}, P04024 (2013).  arXiv:1302.3210 [physics.ins-det].  DOI:10.1088/1748-0221/8/04/P04024.
\bibitem{Gehman:2013B}  V.M.~Gehman.  ``WLS R\&D for the Detection of Noble Gas Scintillation at LBL: seeing the light from neutrinos, to dark matter, to double beta decay''  {\it J.\ Inst.} {\bf 8}, C09007 (2013).  arXiv:1308.2699 [physics.ins-det].
\bibitem{McPherson} McPherson, Inc. (2015) http://www.mcphersoninc.com/
\bibitem{MATLAB} MathWorks, Inc. (2015) \\http://www.mathworks.com/products/matlab/index-b.html
\bibitem{SciPy} The SciPy developers.  (2015) http://www.scipy.org/
\bibitem{pyROOT} The ROOT Team. (2015) https://root.cern.ch/drupal/content/pyroot
\bibitem{Neves2009} F.~Neves, {\em et~al.}, ``Calibration of photomultiplier arrays'',  {\it Astroparticle Physics} {\bf 33} (2009) 15.
\end{thebibliography}
\end{document}